\documentclass[preprint,showpacs,preprintnumbers,amsmath,amssymb]{revtex4}
\include{graphicx}
\graphicspath{{figures/}}

\begin{document}

\title{Ewald Sums for One Dimension}

\author{Bruce N. Miller}

\affiliation{Department of Physics and Astronomy, Texas Christian University,
\\
 Fort Worth, Texas 76129}

\email{b.miller@tcu.edu}

\homepage{http://personal.tcu.edu/ bmiller}

\author{Jean-Louis Rouet}

\affiliation{Institut des Sciences de la Terre d'Orléans - UMR 6113 CNRS/Université
d'Orléans, OSUC, 1A, rue de la Férollerie, F--45071 Orléans Cedex
2, France}
\begin{abstract}
We derive analytic solutions for the potential and field in a one-dimensional
system of masses or charges with periodic boundary conditions, in
other words Ewald sums for one dimension. We also provide a set of
tools for exploring the system evolution and show that it's possible
to construct an efficient algorithm for carrying out simulations.
In the cosmological setting we show that two approaches for satisfying
periodic boundary conditions, one overly specified and the other completely
general, provide a nearly identical clustering evolution until the
number of clusters becomes small, at which time the influence of any
size-dependent boundary cannot be ignored. Finally we compare the
results with other recent work with the hope of providing clarification
over differences these issues have induced. We explain that modern
formulations of physics require a well defined potential which is
not available if the forces are screened directly.
\end{abstract}


\maketitle

\section{Introduction}

One-dimensional models play an important role in physics. While they
are of intrinsic interest, they also provide important insights into
higher-dimensional systems. One-dimensional plasma and gravitational
systems were the first N-body systems simulated with early computers
\cite{lieb_matt,mattis,CL1,CL2}. Plasma systems were used to investigate
Debye screening and thermodynamic equilibrium. In contrast, although
the force law is similar, it was found that the evolution of gravitational
systems toward equilibrium is extremely slow, and still is not completely
explained \cite{WMS,yawn3,tsuchiya4,YM2ma,Joyce_relax}. Recently
spin-offs of these models that, in some cases, are more amenable to
computer simulation have been studied in great depth \cite{posch06,rufforev}.
Three dimensional dynamical simulations are an important component
of modern cosmology. Starting from the precise initial conditions
provided by observations of the cosmic background radiation \cite{WMAP5yr},
various model predictions can be compared with what we observe in
{}``today's universe'' \cite{Virgo}. It was shown by Rouet and
Feix that, in common with the observed positions of galaxies, a cosmological
version of the one-dimensional gravitational system exhibits hierarchical
clustering and fractal-like behavior \cite{Rouet1,Rouet2}. Since
then we, as well as others, have pursued different one-dimensional
cosmological models, finding important attributes such as power law
behavior of both the density fluctuation power spectra and two-body
correlation function, in addition to the influence of dark energy
\cite{Tat,Gouda_powsp,Ricker1d,Joyce_1d,MRGexp,MR_jstat}. In specific
applications, for both plasma and gravity, it is preferable to adopt
periodic boundary conditions because they avoid special treatment
of the boundary and therefore best mimic a segment of the extended
system \cite{Bert_rev,HockneyEastwood,hern_ewald,Virgo} . However,
since they act as a low-pass filter, no information supported on wavelengths
larger than the system size is available. Therefore, in the context
of plasma and gravitational systems, as well as in the Vlasov limit
\cite{braunhepp}, it is necessary to have a sufficiently large system
that contains many Jeans' (or Debye) lengths \cite{peebles2}.

In both plasma and gravitational physics, a large class of one-dimensional
models are defined by a potential energy that satisfies Poisson's
equation. In three dimensions they are represented by embedded systems
of parallel sheets of mass or electric charge that are of infinite
extent. From Gauss' Law, the field from such an element is directed
perpendicular to the surface and has a constant value, independent
of the distance, and proportional to the mass or charge density (per
unit area). In a linear array of sheets of equal mass or charge density,
the force on a given sheet is then simply proportional to the difference
between the number of sheets on the left and right. Problems with
this formulation arise in specific applications where it is necessary
to assume periodic boundary conditions, where the motion takes place
on a torus. A particular case arises in constructing a 1+1 dimensional
model of the expanding universe. In the cosmological setting, astrophysicists
consider a segment of the universe that is following the average expansion
rate. They assume it is large enough to contain many clusters, but
small enough that Newtonian dynamics is adequate to describe the evolution
\cite{Newtap}. Comoving coordinates, in which the average density
remains fixed, are employed, and it is assumed that the system obeys
periodic boundary conditions \cite{HockneyEastwood,Bert_rev}. In
comoving coordinates fictitious forces appear, analogous to the Coriolis
force in a rotating system. The source of the apparent gravitational
field arises from the difference between the actual matter distribution
and a negative background density. To compute the force on a particle
in three-dimensional systems it is possible to carry out {}``Ewald''
sums over the positions of all the other particles in both the system
and all its periodic {}``replicas'' \cite{hern_ewald,brush}. In
one dimension, because the field induced by a particle (mass or charge
sheet) is constant, the solution is not obvious. At first glance it
appears that the force is due to the difference between infinities
and, at second glance, in the cosmological setting, that the negative
background should exactly cancel with the force due to the particles
in each replica. Thus the problem is fraught with ambiguity. To get
a different twist, consider the fact that, in a periodic system, we
are describing motion on a torus, so there is no clear distinction
between left and right. In order to provide conclusive solutions to
these problems, in the following pages we will follow the approach
used by Kiessling in showing that the {}``so-called'' Jeans swindle
is, in fact, legitimate and not a swindle at all \cite{kies_swin}.
For clarity we will focus on the gravitational example.

\section{Periodic Boundary Conditions\label{sec:PBC}}

Let $\rho(x)$ be a periodic mass density (mass density per unit length)
of the one-dimensional system with period $2L$ so that $\varrho(x+2L)=\rho(x).$
In addition write $\rho(x)=\rho_{0}+\sigma(x)$ where $\varrho_{0}$
represents the average of $\rho(x)$ over one period and $\sigma(x)$
is the periodic fluctuation. If, instead, the total mass were bounded,
we could compute the potential in the usual way. The potential function
for a unit mass located at $x'$ is simply $2\pi G\left|x-x'\right|$
so we would normally write\begin{equation}
\Phi(x)=2\pi G\int\left|x-x'\right|\rho(x')dx'\end{equation}
where the integration is over the whole line. Clearly, in the present
situation, this won't converge. Following Kiessling \cite{kies_swin}
we introduce the screening function $\exp\left(-\kappa\left|x-x'\right|\right)$
and define $\Psi$ by\begin{equation}
\Psi(x,\kappa)=2\pi G\int\left|x-x'\right|\exp\left(-\kappa\left|x-x'\right|\right)\rho(x')dx'.\label{eq:scrpot}\end{equation}
 For mass densities of physical interest, for example for both bounded
functions as well as delta functions, $\Psi$ clearly exists in the
current case so long as $\kappa>0$ . The contribution from the average,
or background, term is quickly determined:\begin{equation}
\Psi_{0}=2\pi G(-\frac{\partial}{\partial\kappa})\int\exp\left(-\kappa\left|x-x'\right|\right)\rho_{0}dx'=4\pi G\rho_{0}/\kappa^{2}\end{equation}
and is independent of position. Therefore, although it blows up in
the limit $\kappa\rightarrow0,$ since $\rho_{0}$ is translation
invariant, $\Psi_{0}$ makes no contribution to the gravitational
field. On the other hand we can formulate the contribution to $\Psi$
from $\sigma(x)$, say $\Psi_{\sigma},$ and show that it is well
behaved in this limit.

\subsection{Fourier Representation of the Potential and Field}

Since the density fluctuation $\sigma$ is a mass-neutral periodic
function, we may represent $\sigma(x)$ as a Fourier series\begin{equation}
\sigma(x)={\textstyle \sum_{n}^{'}c_{n}\exp(i\pi nx/L)}\end{equation}
 where the prime indicates that there is no contribution from $n=0.$
Inserting into Eq. (\ref{eq:scrpot}) we find\begin{equation}
\Psi_{\sigma}(x,\kappa)=2\pi G\sum{}_{n}^{'}c_{n}b_{n}\exp(i\pi nx/L)\end{equation}
 where

\[
b_{n}=\int\left|x-x'\right|\exp\left(-\kappa\left|x-x'\right|\right)\exp(in\pi(x'-x)/L)dx'\]
 \[
=\int\left|u\right|\exp\left(-\kappa\left|u\right|\right)\exp(in\pi u)/L)du\]
 \[
=\int\left|u\right|\exp\left(-\kappa\left|u\right|\right)cos(n\pi u/L)du\]
 \[
=2\int_{0}^{\infty}u\exp\left(-\kappa u\right)cos(n\pi u/L)du=2\frac{\kappa^{2}-(\pi n/L)^{2}}{[\kappa^{2}+(\pi n/L)^{2}]^{2}}.\]
 Now, taking the limit $\kappa\rightarrow0$, we find a general expression
for the periodic potential $\phi(x),$\begin{equation}
\phi(x)=-4\pi G{\textstyle \sum_{n}^{'}c_{n}(L/\pi n)^{2}\exp(i\pi nx/L)}\label{eq:perpot}\end{equation}
 and, for the gravitational field $E(x)$,\begin{equation}
E(x)=-\frac{\partial\phi}{\partial x}=4\pi Gi{\textstyle \sum_{n}^{'}c_{n}(L/\pi n)\exp(i\pi nx/L)}.\label{eq:perfield}\end{equation}
Thus, for a periodic distribution of mass, the gravitational field
is well defined. We can think of it as arising from both the mass
in the primitive cell $(-L<x<L)$ and the contribution from the infinite
set of replicas or images. We see immediately that it is a solution
of the Poisson equation, from which the result can also be obtained
more directly from linear independence. For the sake of comparison,
it is worth calculating the field contributed by the primitive cell
alone. This is simply proportional to the difference in the mass on
the right and left of the position $x$:\begin{equation}
E_{p}(x)=2\pi G\int_{-L}^{L}\sigma(x')[\Theta(x'-x)-\Theta(x-x')]\end{equation}
 where $\Theta$ is the usual step function. Substituting for $\sigma(x)$
we find\begin{equation}
E_{p}(x)=4\pi Gi\sum{}_{n}^{'}c_{n}(L/\pi n)[\exp(i\pi nx/L)-(-1)^{n}].\end{equation}
While $E_{p}(x)$ and $E(x)$ are very similar, there is an important
difference: the former is forced to vanish at the endpoints, $x=\pm L$
for all allowed mass distributions within the primitive cell. This
result is expected since we only take into account the field of a
neutral slice. Therefore, in the general case, $E_{p}(x)$ cannot
represent the field on a circle (1-torus).

So far our treatment is quite general and applies to any one-dimensional
periodic mass (or charge) distribution. To gain further insight and
make contact with recent work, let's focus on the situation where
the sources are $2N$ discrete equal-mass points (sheets) with positions
$x_{j}$ that live on the torus with the coordinate boundary points
at $x=L$ and $-L$ identified. Then, in the primitive cell, the density
fluctuation is \begin{equation}
\sigma_{p}(x)=m\sum_{j=1}^{2N}\left[\delta(x-x_{j})-\frac{1}{2L}\right]\label{eq:dis}\end{equation}
 from which we may easily calculate the Fourier coefficients,

\begin{equation}
c_{n}=\frac{1}{2L}\int_{-L}^{L}\exp\left(-i\pi nx'/L\right)\sigma_{p}(x')dx'=\frac{m}{2L}\sum_{j=1}^{2N}\exp\left(-i\pi nx_{j}/L\right),\label{eq:fouco}\end{equation}
for $n$$\neq0.$ Then the gravitational potential and field consists
of the contribution from the primitive cell and all the replicas.
From Eqs. (\ref{eq:perpot},\ref{eq:perfield}) they reduce to\begin{equation}
\phi(x)=-4\pi mGL\sum_{j=1}^{2N}\sum_{n=1}^{\infty}{\textstyle (1/\pi n)^{2}cos(\pi n(x-x_{j})/L),}\label{eq:dispot}\end{equation}

\begin{equation}
E(x)=-\frac{\partial\phi}{\partial x}=-4mG\sum_{j=1}^{2N}\sum_{n=1}^{\infty}{\textstyle (1/n)sin(\pi n(x-x_{j})/L).}\label{eq:disfield}\end{equation}
Thus the Fourier representations of the periodic potential and field
are straightforward.

\subsection{Direct Summation over Replicas\label{sub:PBCsum}}

In the case of three dimensions one cannot do much better than Eqs.(\ref{eq:dispot},\ref{eq:disfield})
since the Ewald sums cannot be represented as simple analytic functions
\cite{brush}. Fortunately, in the present case, we can improve on
this situation. Consider a single particle of mass $m$ located at
$x_{1}$. By summing over replicas, we can compute its contribution
to the screened potential $\Psi(x,\kappa)$ directly:\begin{equation}
\Psi(x,\kappa)=2\pi mG\sum_{r=-\infty}^{\infty}\left|x-x_{1}-2rL\right|exp\left(-\kappa\left|x-x_{1}-2rL\right|\right)\end{equation}
 \begin{equation}
=2\pi mG(-\frac{\partial}{\partial\kappa})\sum_{r=-\infty}^{\infty}exp\left(-\kappa\left|y_{1}-2rL\right|\right)\end{equation}
 \begin{equation}
=2\pi mG(-\frac{\partial}{\partial\kappa})\sum_{r=-\infty}^{\infty}\{exp\left(-\kappa(y_{1}-2rL)\right)\Theta\left(y_{1}-2rL\right)+exp\left(\kappa(y_{1}-2rL)\right)\Theta\left(-y_{1}+2rL\right)\}\label{eq:dbsum}\end{equation}
where $y_{1}=x-x_{1}$ . Choose integers $r_{<}(y_{1})$and $r_{>}(y_{1})$
such that $r_{<}\leq y_{1}/2L\leq r_{>}=r_{<}+1$, i.e. $y_{1}/2L$
is bounded from below and above by this pair of adjacent integers.
Then\begin{equation}
\sum_{r=-\infty}^{\infty}exp\left(-\kappa(y_{1}-2rL)\right)\Theta\left(y_{1}-2rL\right)=exp\left(-\kappa(y_{1}-2r_{<}L)\right)\sum_{s=-\infty}^{0}exp(\kappa2Ls)\end{equation}
and similarly for the second sum in Eq.(\ref{eq:dbsum}). Therefore
each of the sums in Eq.(\ref{eq:dbsum}) can be evaluated in terms
of a geometric series to obtain the screened potential \begin{equation}
\Psi_{\sigma}(x,\kappa)=2\pi mG(-\frac{\partial}{\partial\kappa})\left\{ \left[exp\left(-\kappa Y_{<}\right)+exp\left(+\kappa Y_{>}\right)\right]/\left(1-\exp\left(-2\kappa L\right)\right)-1/\kappa L\right\} \end{equation}
where $Y_{<}=y_{1}-2r_{<}L$ , etc., and we have subtracted the contribution
from the average or background density, $m/2L$. Evaluating the derivative
and then taking the limit $\kappa\rightarrow0$, we obtain the gravitational
potential $\phi_{1}$ due to a single particle at $x_{1}$:\begin{equation}
\phi_{1}\left(x\right)=-\frac{\pi mG}{2L}\left(Y_{>}^{2}+Y_{<}^{2}\right).\label{eq:sinparpot}\end{equation}

It is important to recognize that $\phi_{1}\left(x\right)$ is a periodic
function of its argument and can be evaluated anywhere on the periodic
extension of the torus, i.e. on the real line. As physicists we are
typically interested in values of $x$ and $x_{1}$ in the primitive
cell, i.e. for $-L\leq x,\: x_{1}<L$ with the points at $x=\pm L$
identified. Then we quickly find that, for $y_{1}\geq0,$ $0\leq y_{1}/2L<1$
whereas for $y_{1}<0,$ $y_{1}/2L$ is sandwiched between $[-1,0)$
. Either way, the potential $\phi_{1}$, and therefore the field $E_{1}$,
can be represented as \begin{equation}
\phi_{1}\left(x\right)=2\pi mG\left[\left|x-x_{1}\right|-\frac{1}{2L}\left(x-x_{1}\right)^{2}\right]\label{eq:sinpartpotfin}\end{equation}
 \begin{equation}
E_{1}(x)=-\frac{\partial\phi_{1}}{\partial x}=2\pi mG\left[\frac{1}{L}(x-x_{1})+\Theta(x_{1}-x)-\Theta(x-x_{1})\right].\label{eq:sinpartfield}\end{equation}
 Thus, in addition to the direct contribution from the mass located
at $x_{1}$, there is an additional quadratic term in the potential
and linear term in the field. Although these contributions are simple,
care must be taken in their interpretation. They are not, respectively,
equal to the potential and field contributed by the component of the
background located between the point of application $x$ and the location
of the source $x_{1}$, but rather twice as large!

A number of observations are in order. First of all, it is obvious
that these functions reproduce exactly the Fourier series derived
above for the case of a single particle. Second, in the limit
$L\rightarrow\infty,$ they reduce to the familiar results on the
line. Third, they are strictly functions of the displacement
$x-x_{1}.$ This is important as all points on the torus are
equivalent: there are no special positions or intervals. Fourth, it
is not necessary to distinguish in which direction the distance
between the points $x$ and $x_{1}$ is measured. Going in either
direction around the torus yields the same value of $\phi_{1}.$
Fifth, defining $\Theta(0)=\frac{1}{2}$, the field vanishes at both
$x=x_{1}$ and at $x=x_{1}-L$ mod$\left(2L\right)$, i.e. half way
around the torus. Finally, when $x_{1}$ traverses the point at $L$
and reappears at $-L$, or vice-versa, there is no change in the
field at $x$ as we would expect from the physics.

\subsection{Symmetry-Based Derivation}

In the above we have employed the machinery of a screening function
to obtain the desired result. While it has all the right properties,
it is worth asking if we could have obtained it from a simpler route.
We seek a solution of Poisson's equation for a single mass located
at $x_{1}$ ,

\begin{equation}
\frac{\partial^{2}\phi}{\partial x^{2}}=4\pi G\sigma(x),\label{eq:poiseq}\end{equation}
 where, in the primitive cell, \begin{equation}
\sigma=\sigma_{p}(x)=m\left[\delta\left(x-x_{1}\right)-\left(\frac{1}{2L}\right)\right].\label{eq:den1}\end{equation}
 The general solution is \begin{equation}
\phi(x)=2\pi mG\left[\left|y_{1}\right|-\frac{1}{2L}y_{1}^{2}+by_{1}\right],\end{equation}
 yielding \begin{equation}
E(x)=-\frac{\partial\phi}{\partial x}=2\pi
mG\left[\frac{1}{L}(y_{1})-\Theta(y_{1})+\Theta(-y_{1})+b\right]\end{equation}
where $b$ is an arbitrary constant, as before $y_{1}=x-x_{1},$ and
we have chosen the additive constant in the expression for $\phi$
such that $\phi(x=x_{1})=0$. Symmetry requires that there is no
preferred direction on the torus. Therefore, regardless of the
location of the mass at $x_{1}$, the average of the field in
$[-L,L)$ must vanish. Then we immediately obtain $b=0$ and the
results given in subsection \ref{sub:PBCsum} above. We could also
have arrived at this conclusion by noting that, for the same reason,
$\phi(x)$ can only depend on the distance between $x$ and $x_{1}$.
Finally, the requirement that $\phi(L)=\phi(-L)$ also demands the
same conclusion. The fact that our limiting procedure, which is
based on an exponential screening function, leads to a unique
solution of the Poisson equation increases our confidence that the
choice of a different screening function, e.g. a Gaussian, would not
result in a different potential or field.

For the sake of comparison, and to understand the connection with
other recent work, it's worthwhile to consider $E_{1p}(x)$, the field
generated solely by $\sigma_{p}$, the charge distribution in the
primitive cell. This would be the correct field if the net contribution
from each replica vanished. We quickly find \begin{equation}
E_{1p}(x)=2\pi mG\left[\frac{1}{L}x+\Theta(x_{1}-x)-\Theta(x-x_{1})\right].\label{eq:E1prim}\end{equation}
Among other problems, note that it does not satisfy the symmetry requirements
discussed above. We will return to this formulation in the ensuing
discussion.

\section{N-body Simulation}

In carrying out a simulation, we need to know the field acting on
each particle. Summing over all the contributions, the total field
at $x$ arising from the complete system of particles is then simply

\begin{equation}
E(x)=4\pi mG\left[\frac{N}{L}(x-x_{c})+\frac{1}{2}(N_{R}(x)-N_{L}(x))\right]\label{eq:totfield}\end{equation}
where, in Eq. (\ref{eq:totfield}), $x_{c}$ is the center of mass
of the $2N$ particles in $[-L,\: L)$ and $N_{R}(x),\, N_{L}(x)$
are the number of particles to the right (left) of $x$ counted on
the segment $[-L,\, L).$ Since, from Eq. (\ref{eq:sinpartfield}),
the field from a single particle vanishes at its location, Eq. (\ref{eq:totfield})
gives the correct field acting on each particle, i.e. $E_{j}=E(x=x_{j}).$
The presence of the center of mass in Eq. (\ref{eq:totfield}) means
that the instantaneous field experienced by each particle depends
on the dipole moment of the system. The dependence on the center of
mass is essential as it insures that when a particle passes from $x=L$
to $x=-L$ or vice-versa, there is no change in the field experienced
by each particle. This was recognized as a basic problem in simulations
of the one-dimensional, single-component plasma some time ago (see
\cite{eldfeix} reprinted in \cite{lieb_matt}). Perhaps the correct
mathematical form of the field was not known. In order to avoid discontinuous
jumps in the field, a polarization charge was artificially induced
at the system boundaries, and was changed whenever a particle {}``switched
sides''. In this way the boundaries were seen as initially neutral
reservoirs of particles. When a system particle enters one reservoir,
another particle escapes from the other, so the boundaries are no
longer neutral.

For systems of interest the equations of motion of the system of particles
can frequently be cast in the form \begin{equation}
\frac{dx_{j}}{dt}\equiv v_{j},\;\frac{dv_{j}}{dt}+\gamma v_{j}=E(x_{j})\label{eq:eqmo}\end{equation}
where the value of the friction constant $\gamma$ depends on the
particular model \cite{MRGexp,MR_jstat}. In the cosmological setting
the time has been rescaled to retain the simplicity of the equations
of motion and increases exponentially with the comoving time coordinate
\cite{MRexp,MR_jstat}. By carefully summing over the index $j$ we
find that the velocity of the center of mass obeys the simple equation\begin{equation}
\frac{dx_{c}}{dt}\equiv v_{c}\,,\quad\frac{dv_{c}}{dt}+\gamma v_{c}=0.\label{eq:com}\end{equation}
When a particle traverses the coordinate boundary at $x=L$ the center
of mass changes discontinuously. However, the center of mass velocity
is a smooth function of time so the first order equation for $v_{c}(t)$
can be integrated immediately: $v_{c}(t)=v_{c}(0)\exp(-\gamma t)$.
In particular, for the special case where the center of mass is initially
at rest, its velocity maintains its initial value. On the other hand,
$x_{c}(t)$ will change abruptly with each boundary traversal.

In carrying out a simulation it is necessary to obtain the crossing
times of adjacent particles. Starting at $x=-L$, label the particles
according to their order so that $x_{2N}>\cdots>x_{j+1}>x_{j}>\cdots>x_{1}$
and define $z_{j}=x_{j+1}-x_{j}$ , the displacement between the adjacent
particles. Then, from Eqs. (\ref{eq:totfield}, \ref{eq:eqmo}), we
find that $z_{j}$ obeys \begin{equation}
w_{j}\equiv\frac{dz_{j}}{dt},\,\quad\frac{dw_{j}}{dt}+\gamma w_{j}=4\pi mG\left(\frac{N}{L}z_{j}-1\right)\label{eq:zmo}\end{equation}
 for $j=1,\ldots,$$\:2N-1.$ To complete the ring we continue in
the same sense and determine the rate of change of the displacement
between $x_{2N}$ and $x_{1}$, that is $2L+x_{1}-x_{2N}\equiv z_{2N}$
, and find that it too conveniently satisfies Eq.(\ref{eq:zmo}).
Thus, by defining a positive direction, or orientation, on the torus,
we can keep track of the all the relative positions between nearest-neighbor
particle pairs subject to the constraints \begin{equation}
\sum_{1}^{2N}z_{j}=2L,\:\sum_{1}^{2N}w_{j}=0.\label{eq:constraints}\end{equation}
 Since, from Eq.(\ref{eq:com}) we already know the velocity of the
center of mass $v_{c}$, we can invert the set $\left\{ w_{j}\right\} $,
$v_{c}$, to obtain the particle velocities $v_{j}$ at any time using
a matrix inversion given by Rybicki \cite{Rybicki}. Conceptually,
since the particles have equal mass, except for labels, they appear
to experience elastic collisions with their neighbors. When the positions
of a pair of particles cross, the particles exchange accelerations,
but the velocities are continuous. At such an event the labels of
the particle pair are exchanged to maintain the ordering in the given
direction. As time progresses we see that, for this completely ordered
system, when $z_{j}=0,$ i.e. when the $j^{th}$ and $j+1^{st}$ particle
cross, $w_{j}$ changes sign. Moreover, since the particle labels
have been exchanged, the velocities of the two neighboring pairs,
$z_{j-1}$ and $z_{j+1}$ , also change discontinuously, i.e., $w_{j-1}\rightarrow w_{j-1}+w_{j}$
and $w_{j+1}\rightarrow w_{j+1}+w_{j}$ at the crossing time. This
is all the information required to carry out a simulation. While we
have focused on the gravitational system of equal masses, it's straightforward
to extend the approach to the case of unequal masses, as well as to
the single and two-component plasmas.

\begin{figure}[ht]
\includegraphics[angle=-90,width=0.475\textwidth]{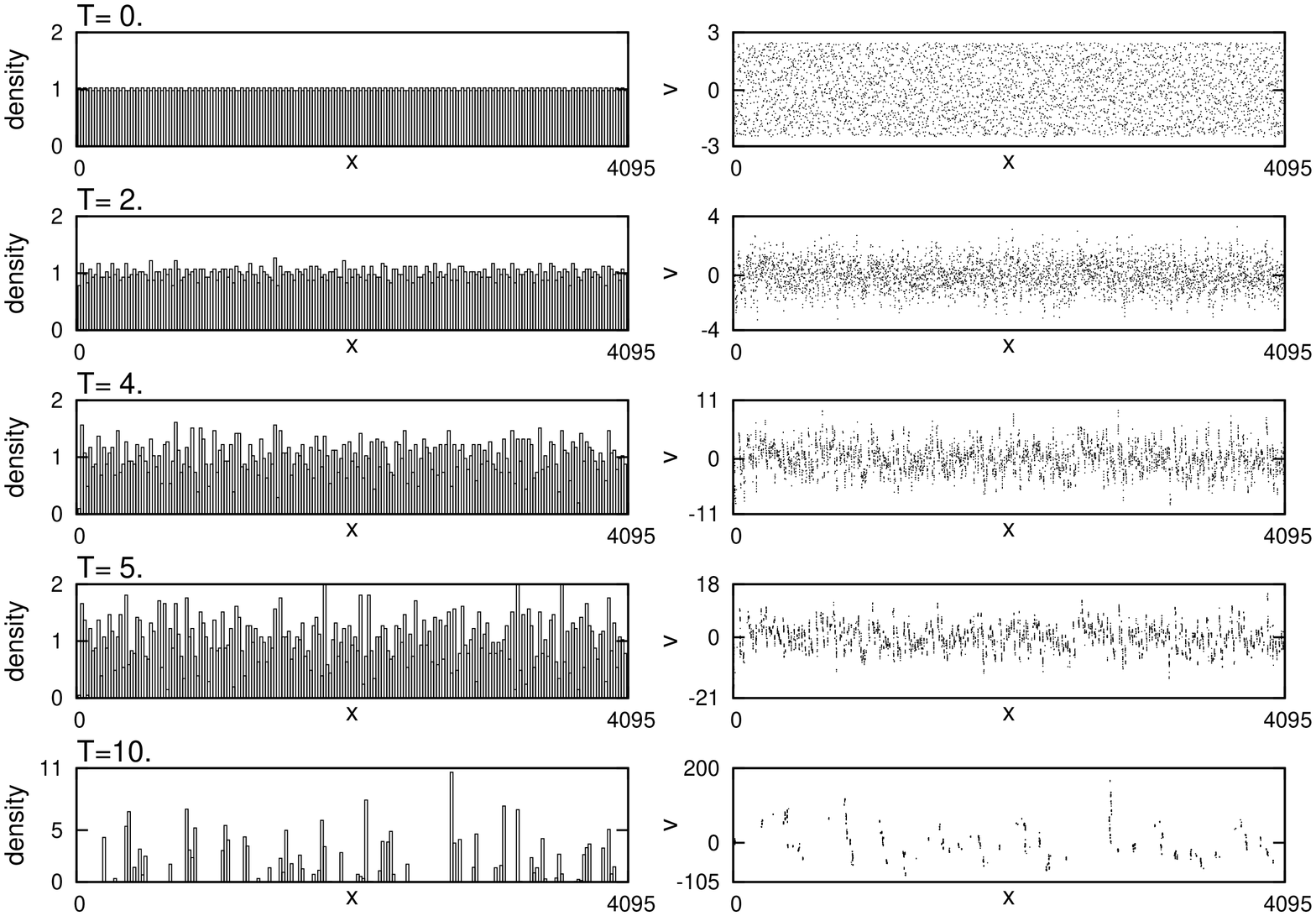}
\includegraphics[angle=-90,width=0.475\textwidth]{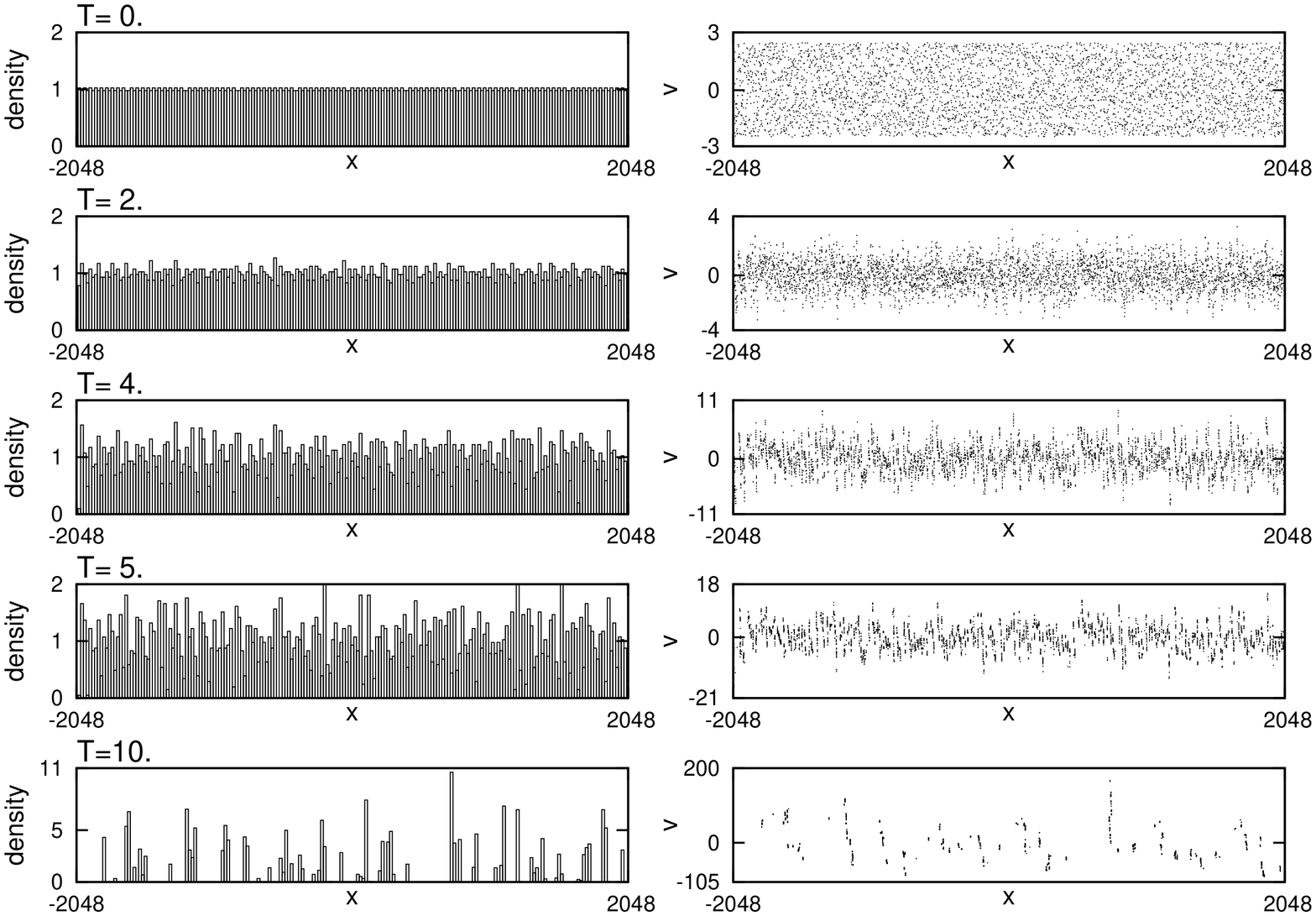}
\includegraphics[angle=-90,width=0.475\textwidth]{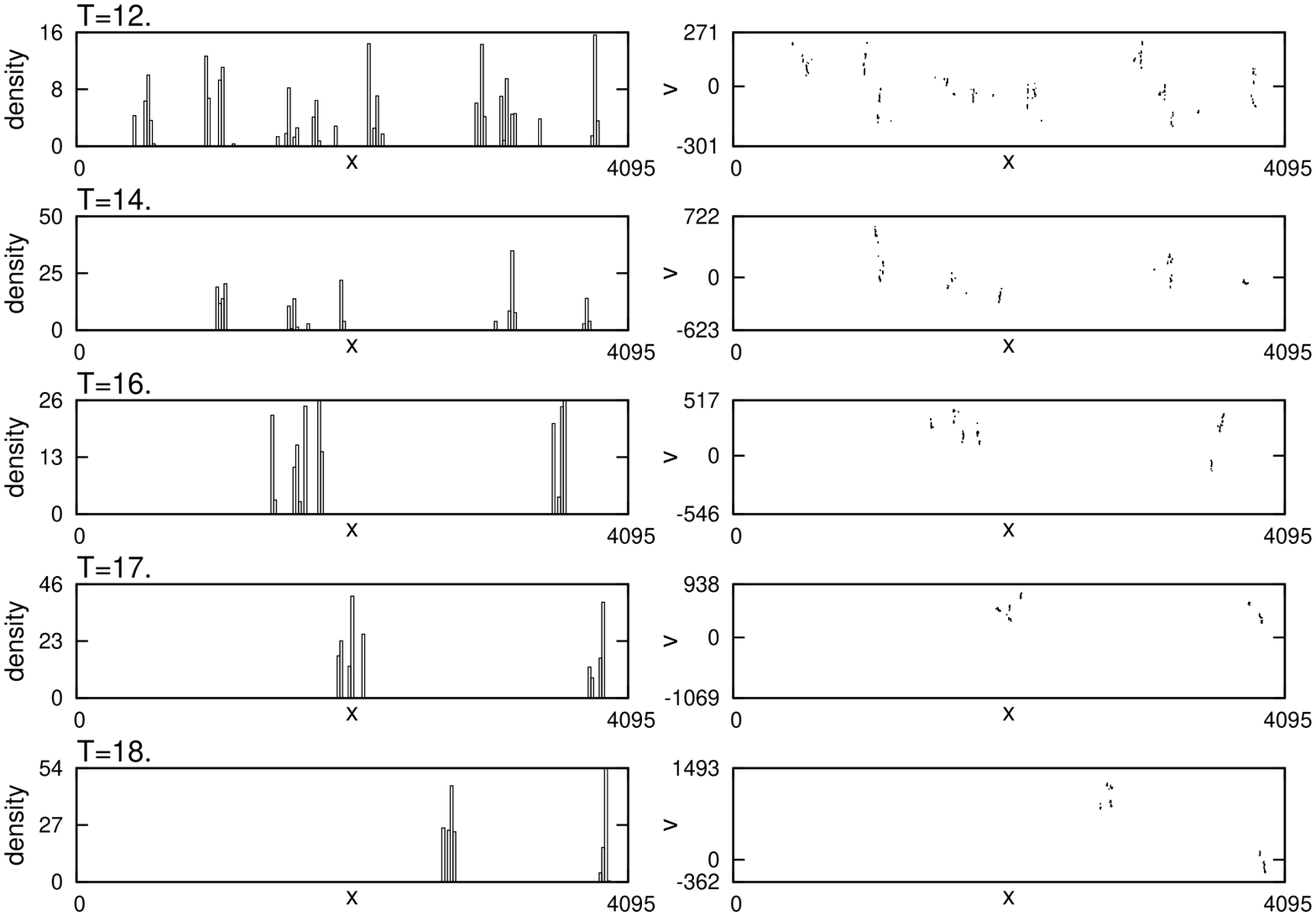}
\includegraphics[angle=-90,width=0.475\textwidth]{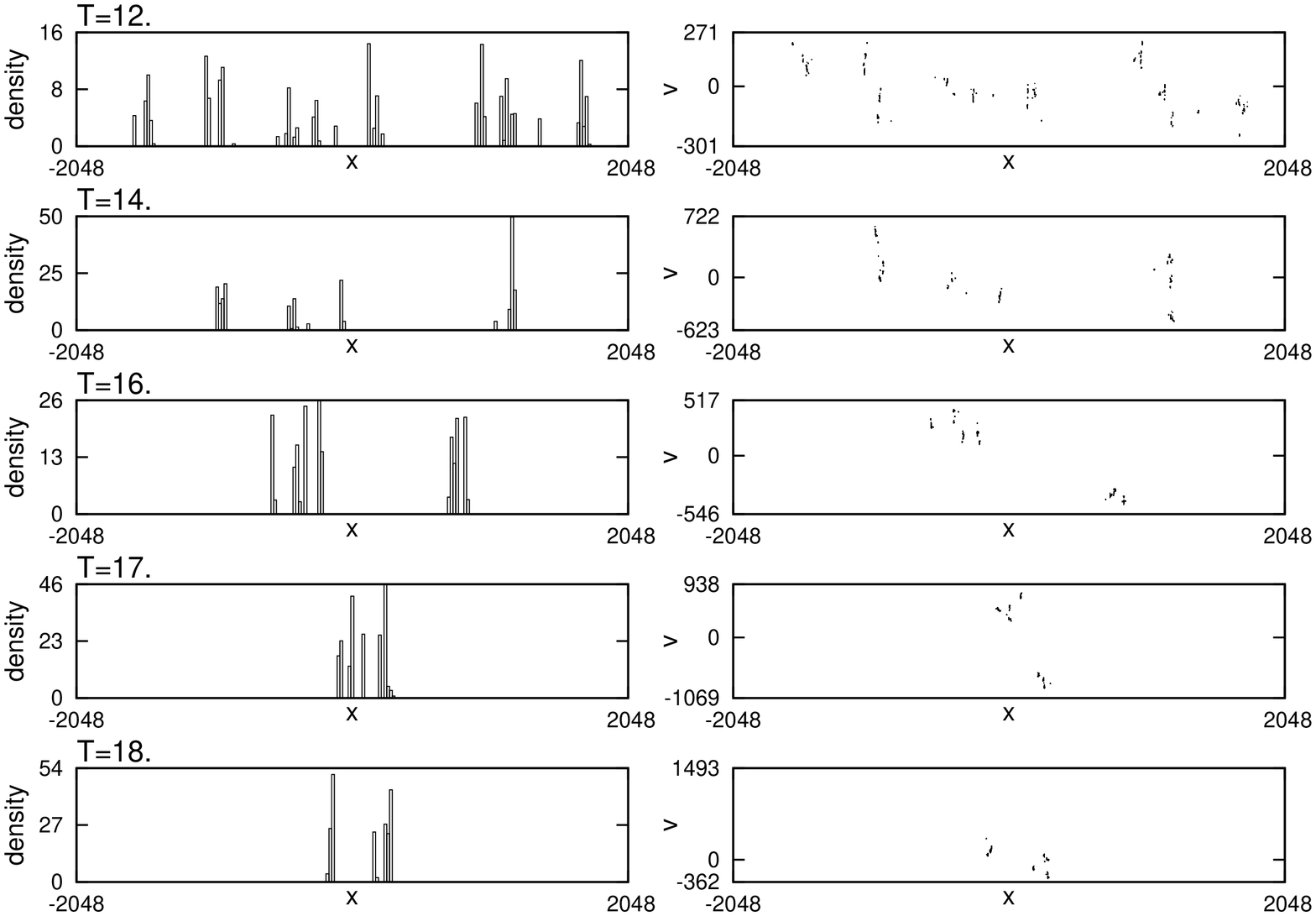}
\caption{\label{aa}Snapshots of the density and phase space at
different times for a symmetric system(left) and a periodic system
(right). The initial positions and velocities of all particles are
the same for both simulations.} \end{figure}

Below, in Fig. \ref{aa}, we present a series of snapshots from two
recent simulations of a one-dimensional model of the expanding universe
in comoving coordinates in which only gravitational forces apply.
The model used was the one introduced originally by Rouet and Feix
\cite{Rouet1,Rouet2}, i.e. Eq. (\ref{eq:eqmo}) with $\gamma=\frac{1}{\sqrt{2}}$.
Each simulation employs $2^{12}-1$ particles, and identical initial
conditions were drawn from a uniform waterbag configuration in the
$\mu$-space. The dimensionless, scaled, unit of time is expressed
in terms of the Jeans' period and the dimensionless length is simply
the number of particles \cite{MRGexp,MR_jstat}. It's important to
keep in mind that the scaled time is an exponentially increasing function
of the comoving time coordinate \cite{MRexp,MR_jstat}. The left sequence
shows the evolution under the symmetric version of the system as it
was originally employed (see the discussion below), while the sequence
on the right exhibits the evolution obtained with Eq.(\ref{eq:totfield}).
In each sequence, the left hand column represents a histogram of positions,
while the right hand column shows the positions in the position-velocity
plane, i.e. what statistical physicists call $\mu$-space and astrophysicists
call phase space.

\begin{figure}[ht]
\centerline{\includegraphics[angle=-90,width=\textwidth]{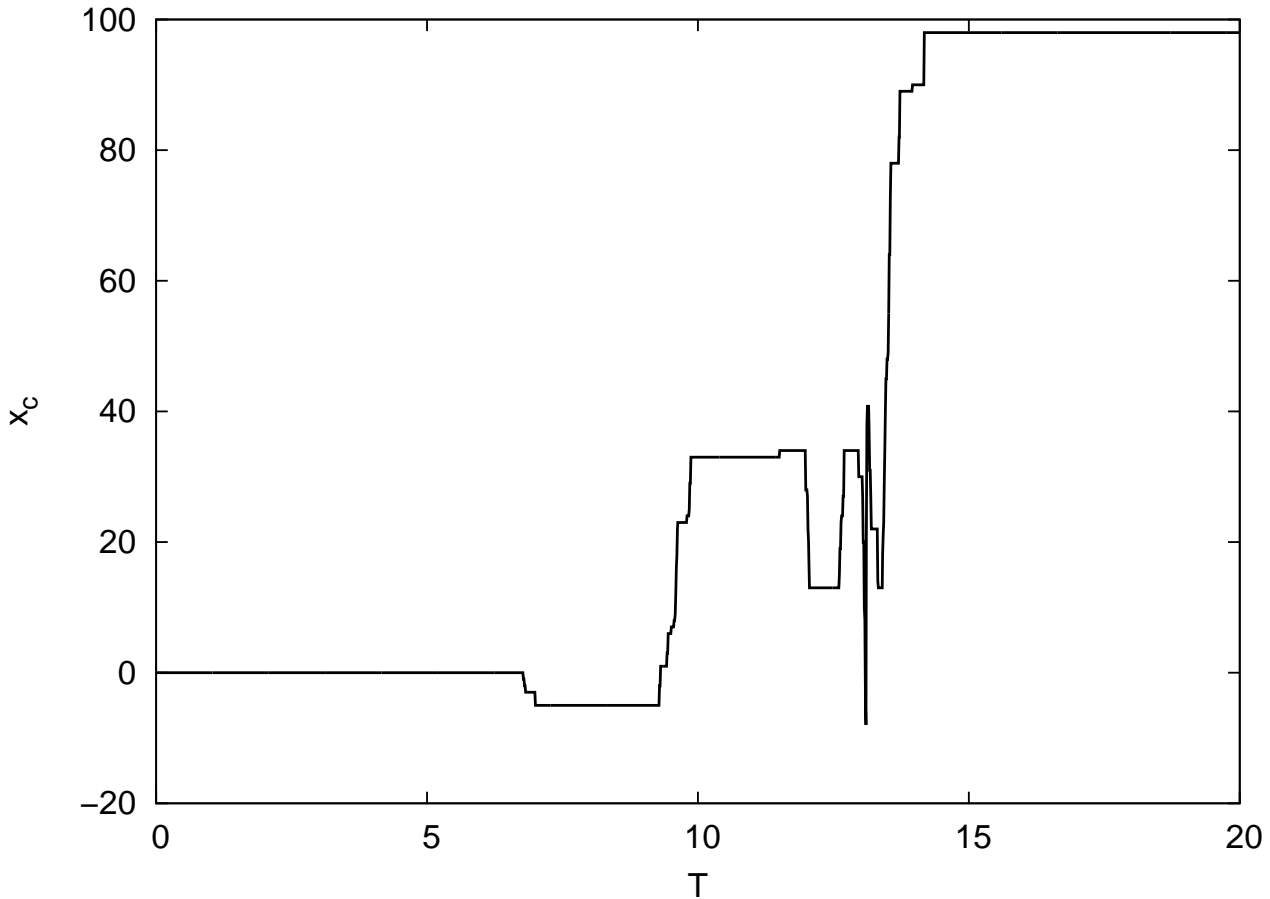}}
\caption{\label{xc} Time evolution of the center of mass for a
simulation containing 255 particles. Each time a particle crosses
the coordinate system boundary there is a corresponding variation of
unity.} \end{figure}

We observe an initial shrinkage in the phase plane due to the friction
constant, followed by the break-up of the system into many small clusters
that results from the intrinsic gravitational instability \cite{peebles2}.
As time progresses, in each case we observe the continual, self-similar
creation of larger clusters from smaller ones. Comparing the two snapshot
sequences, we see that initially, and for some time, they are virtually
identical. Before $T=10$ there is no discernable difference between
the two runs in the location of cluster positions in the phase plane.
Then, at $T\simeq10$ , we do notice a minute difference between them.
Even at $T=12$ they are remarkably similar! Finally, at $T\simeq14$
, we observe a noticeable difference between the runs: in the first
sequence the few remaining clusters are slowly gathering towards the
right hand coordinate boundary whereas, in the second, they remain
evenly dispersed throughout the system. This occurs because there
is a natural bias in the original implementation of the model. By
forcing the system to be symmetric, the motion effectively takes place
in a fixed external potential proportional to $-x^{2}$ (as explained
below, for the symmetric system there is a {}``ghost'' system of
particles for $x<0$ that we don't display). Consequently as the system
loses {}``energy'', matter naturally gathers near the right hand
coordinate boundary, in the neighborhood of the potential minimum.
In the second simulation the potential is translation invariant on
the torus so there are no favored locations and the clusters remain
uniformly dispersed. Notice that, by the time we can see any significant
differences, there are only about five clusters in the system so the
boundaries are starting to play a role in the evolution. Thus the
simulation is no longer representative of the larger system. At earlier
times, while the number of clusters was large, as we would intuitively
anticipate, the boundaries had little effect. To illustrate the working
of the algorithm, in Fig. \ref{xc} we plot the position of the center
of mass for a small system ($255$ particles). We can see how $x_{c}$
shifts in time. There are periods of little change, as well as periods
of significant change, when a cluster or group encounters a coordinate
boundary.

\section{Discussion and Conclusions}

There is a long history of work on one-dimensional plasma models,
both theoretical and computational, dating back to the 1960's\cite{lieb_matt,mattis}.
Some papers explicitly discuss the case of periodic boundary conditions.
As we pointed out earlier, in simulations of the single-component
plasma Eldridge and Feix employed a polarization field at the boundaries
to control the discontinuities in the electric field \cite{eldfeix}.
In a review paper Kunz gives an analytical expression for the potential,
but there is no derivation \cite{kunz}. Periodic boundary conditions
are required for cosmological simulations \cite{HockneyEastwood,Bert_rev,hern_ewald}.
The first one-dimensional cosmological simulations were carried out
by Rouet and Feix \cite{Rouet1,Rouet2}. To avoid the problem of introducing
a {}``polarization'' field at the boundaries, they assumed that
the system was perfectly symmetric at all times, i.e. for every particle
at $x_{j}>0$ with velocity v$_{j}$ there is an image or {}``ghost''
particle at position $-x_{j}$ with velocity $-v_{j}$. When a particle
reaches the coordinate boundary, the image particle is there to meet
it. Thus the 2N-particle system is equivalent to an N-particle system
with $0<x<L$ with reflecting boundaries. Notice that this construction
forces the center-of-mass position and velocity to vanish, simplifying
the equations of motion (see Eq.(\ref{eq:totfield})).

In other one-dimensional simulations, Aurell et al. employed open
boundaries \cite{Aurell,Aurell2}. In their studies an initially localized
fluctuation inter-penetrates a quiescent region. The field they employed
is essentially given by Eq. (\ref{eq:E1prim}). Gouda and Yano \cite{Gouda_powsp},
as well as Tatekawa and Maeda \cite{Tat}, employed the Zeldovich
approximation \cite{peebles2}. Details concerning the type of boundary
conditions employed were not discussed in these works but, in contrast
with Eq. (\ref{eq:totfield}) above, the Zeldovich approximation,
as normally derived, does not depend on the system center of mass.
Gabrielli et al. have studied the behavior of an infinite system of
sheets perturbed from lattice positions \cite{Joyce_1d}. They also
employed the screening function introduced by Kiessling to obtain
an analytical expression for the field so, in spirit, their work is
closely related to ours. However, there is a surprising difference
in the expression they obtained for the gravitational field for the
case of periodic boundary conditions, which is also given by Eq. (\ref{eq:E1prim}).
Since it lacks the explicit dependence on $x_{c}$, it is not translation
invariant on the torus and there is a discontinuity in the field when
a particle passes through a coordinate boundary at $x=\pm L$. Consequently
it doesn't represent true motion on a torus. While it is tempting
to contemplate that the re-introduction of particles that leave from
$x=L$ at $x=-L$ , and vice-versa, is adequate to guarantee periodic
boundary conditions, this is not the case. An additional difficulty
is that the field they present is self-referential, i.e. since $E_{1p}(x=x_{1})\neq0$
(see Eq.(\ref{eq:E1prim})), the field generated by a single particle
will induce an acceleration of itself.

The approach taken by Gabrielli et al. and the one taken here are
remarkably similar, so it's worth trying to sort out why they
produce different results. Here, following Kiessling
\cite{kies_swin}, we have taken the usual approach of screening the
gravitational potential so in Eq.(\ref{eq:scrpot}) we are simply
starting with a one-dimensional version of the Yukawa potential. In
contrast, in \cite{Joyce_1d}, Gabrielli et al. effectively screen
the field of a particle located at $x_{1}$ directly by
$exp(-\kappa\left|x-x_{1}\right|$). It is straightforward to verify
that the potential corresponding to this screened field is $(2\pi
mG/\kappa)[1-exp(-\kappa\left|x-x_{1}\right|)]$ . Then, for an
arbitrary mass distribution $\rho(x),$ the corresponding
{}``screened'' potential is \begin{equation} (2\pi
mG/\kappa)\intop[1-exp\left(-\kappa\left|x-x'\right|\right)]\rho(x')dx'.\label{eq:potGab}\end{equation}
For an extended mass distribution, for example the periodic system
considered here, it is apparent that this won't converge.
Contemporary foundations of physics, both classical and quantum, are
based on a Lagrangian or Hamiltonian formulation in which the
potential plays a more fundamental role than the force. A good
example is Feynman's dissertation where he develops the path
integral \cite{feyn-dis}. There Newton's laws arise from paths for
which the action is an extremum. To extend the current model to,
say, the quantum regime, the availability of a clearly defined
potential, such as Eq.( \ref{eq:sinpartpotfin}), is essential.

In conclusion, we have derived analytic solutions for the potential
and field in a one-dimensional system of masses or charges with periodic
boundary conditions, i.e. Ewald sums for one-dimension. We have seen
that each particle in such a system carries with it its own neutralizing
background, without which the potential energy cannot be defined.
For a system of particles, we have shown that the system {}``polarization''
or center of mass must be explicitly included in the force law. We
have also provided a set of tools for exploring the system evolution
and have shown that it's possible to construct an efficient algorithm
for accomplishing this. In the cosmological setting we have shown
that the difference between the choice of completely symmetric, or
just periodic, boundary conditions plays an insignificant role in
the evolution until the number of clusters becomes small, at which
time the influence of any boundary condition will become important.
Finally, we showed that directly screening the force, as in \cite{Joyce_1d},
instead of the potential leads to a divergent potential function for
an extended system and is therefore not suitable for the preferred
formulations of physics based on variational priniciples. In subsequent
work we will explore other settings where boundaries play a more prominent
role.
\begin{acknowledgments}
The authors benefitted from interactions with Igor Prokhorenkov and
Paul Ricker, the hospitality of the Université d'Orléans, and the
support of the Research Foundation and the division of Technology
Resources at Texas Christian University.
\end{acknowledgments}

\bibliography{gravbib6}

\end{document}